\documentclass[twocolumn,showpacs,superscriptaddress,email,floatfix,longbibliography,pra]{revtex4-2}

\usepackage{amsmath}
\usepackage{inputenc}
\usepackage{graphicx}
\usepackage{braket}
\usepackage{longtable}
\usepackage{bbold}
\usepackage{amsmath,amssymb,amsfonts,bbm,graphicx,hyperref,color}
\usepackage{physics}
\usepackage{multirow}
\usepackage{graphicx}
\usepackage[table,xcdraw]{xcolor}
\usepackage{natbib}
\usepackage{hyperref}
\usepackage{xr}

\begin{document}
\title{Accelerating the approach of dissipative quantum spin systems towards stationarity through global spin rotations}
\author{Simon Kochsiek}
\affiliation{Institut f\"ur Theoretische Physik, Universit\"at Tübingen, Auf der Morgenstelle 14, 72076 T\"ubingen, Germany}
\author{Federico Carollo}
\affiliation{Institut f\"ur Theoretische Physik, Universit\"at Tübingen, Auf der Morgenstelle 14, 72076 T\"ubingen, Germany}
\author{Igor Lesanovsky}
\affiliation{Institut f\"ur Theoretische Physik, Universit\"at Tübingen, Auf der Morgenstelle 14, 72076 T\"ubingen, Germany}
\affiliation{School of Physics and Astronomy and Centre for the Mathematics and Theoretical Physics of Quantum Non-Equilibrium Systems, The University of Nottingham, Nottingham, NG7 2RD, United Kingdom}

\begin{abstract}
We consider open quantum systems whose dynamics is governed by a time-independent Markovian Lindblad Master equation. Such systems approach their stationary state on a timescale that is determined by the spectral gap of the generator of the Master equation dynamics. In the recent paper [Carollo et al., Phys. Rev. Lett. \textbf{127}, 060401 (2021)] it was shown that under certain circumstances it is possible to exponentially accelerate the approach to stationarity by performing a unitary transformation of the initial state. This phenomenon can be regarded as the quantum version of the so-called Mpemba effect. The transformation of the initial state removes its overlap with the dynamical mode of the open system dynamics that possesses the slowest decay rate and thus determines the spectral gap. While this transformation can be exactly constructed in some cases, it is in practice challenging to implement. Here we show that even far simpler transformations constructed by a global unitary spin rotation allow to exponentially speed up relaxation. We demonstrate this using simple dissipative quantum spin systems, which are relevant for current quantum simulation and computation platforms based on trapped atoms and ions.
\end{abstract}

\maketitle

\section{Introduction}
The study of open quantum systems is an important sub-field in modern physics. It considers ensembles of quantum particles that are weakly coupled to an external bath. This coupling gives rise to irreversible processes and a relaxation dynamics that typically leads to a stationary state. In the simplest possible setting this dynamics is governed by a Markovian Lindblad Master equation \cite{lindblad1976,gorini1976,breuer2002}, which evolves the quantum state $\rho$ of the system in time. This approach is applicable to a wide range of phenomena, including simple processes, such as spontaneous decay, or sophisticated many-body effects, such as sub- and superradiance \cite{guerin2016}. Moreover, this formulation allows to describe stationary state phases, transitions among them \cite{kessler2012,casteels2017} and even protocols for (quantum) computation \cite{verstraete2009} and the creation of correlated many-body states \cite{lin2013,carr2013,bardyn2013} which may be relevant in the realm of quantum technologies. 

In the context of these applications one may be interested in approaching the stationary state, i.e. the end result of a computation or a desired correlated state, on a timescale which is as short as possible. In case of Markovian (Lindblad) time-evolutions, this timescale is dictated by the spectrum --- in particular by the spectral gap --- of the dynamical generator. This gap is, however, difficult to alter without changing the characteristics of the stationary state itself. An alternative approach to this problem is offered by the so-called Mpemba effect which was first discussed in the context of classical non-equilibrium problems. Originally, it describes the counterintuitive phenomenon that water, which is initially at a high temperature, freezes faster than it would when starting from a lower temperature. It is named after Erasto B. Mpemba who has discovered this effect while preparing ice-cream,  at that time being a schoolboy in Tanzania \cite{mpemba1969}. The observed accelerated approach to the stationary state turns out to originate from the fact that a high-temperature thermal state has lower overlap with slowly decaying dynamical modes than a thermal state with low temperature \cite{lu2017,lasanta2017,nava2019,baity-jesi2019,gal2020,kumar2020,gijon2019,klich2019,torrente2019,bechhoefer2021}. 

\begin{figure}
    \centering
    \includegraphics[width=\columnwidth]{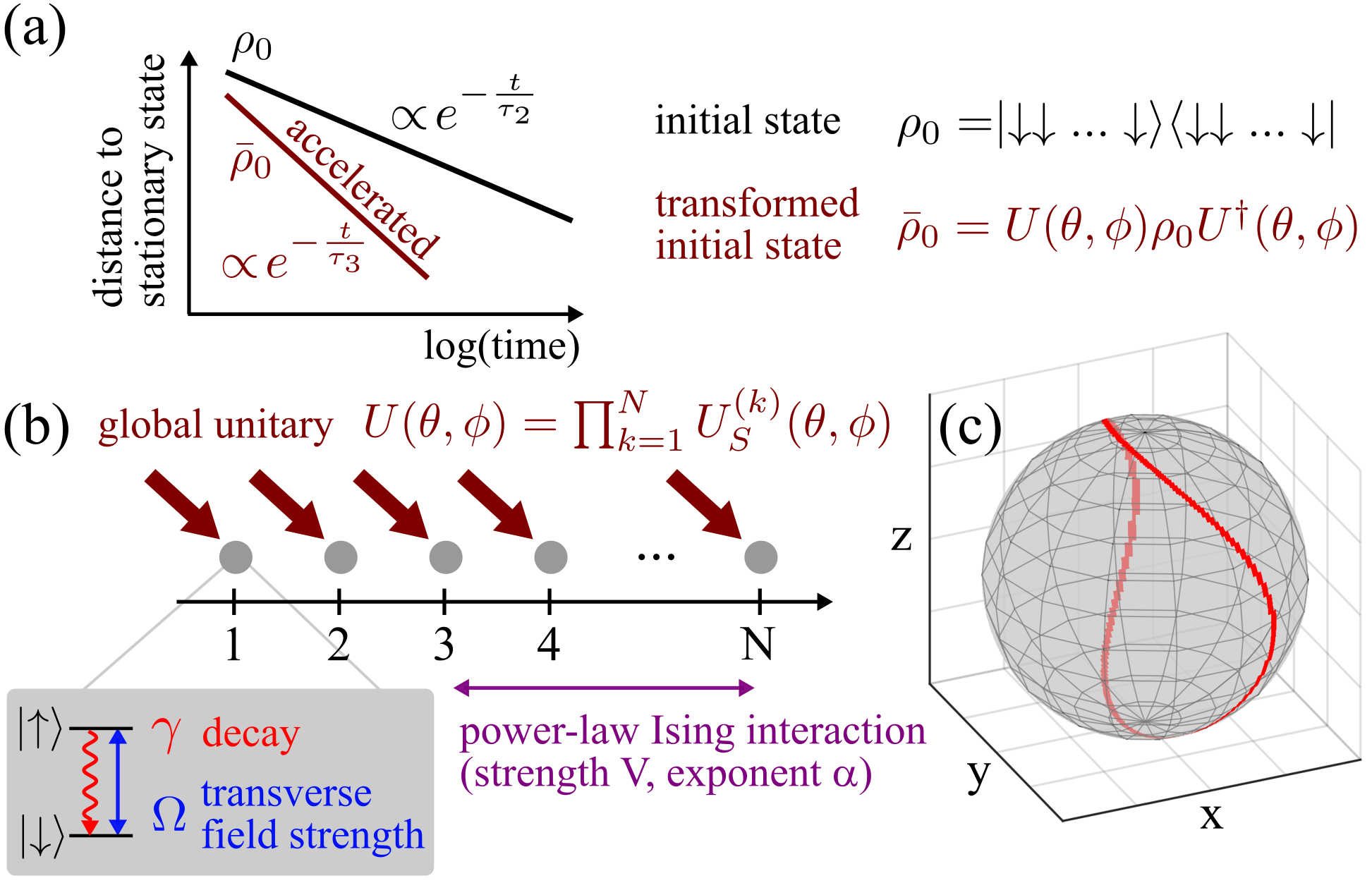}
    \caption{\textbf{Quantum Mpemba effect and open quantum Ising spin chain.} (a) The application of a unitary transformation $U(\theta,\phi)$ to the initial state $\rho_0$ may accelerate the approach of an open quantum system to its stationary state. This is achieved when the unitary removes the overlap of the transformed initial state $\bar{\rho}_0$ with the slowest decaying mode, with lifetime $\tau_2$ (inverse of the negative real part of the spectral gap $\lambda_2$, see main text). In that case the longtime dynamics is governed by the lifetime $\tau_3 < \tau_2$. In this work we consider global unitary spin rotations, which are parametrised by the polar and the azimuthal angles $\theta$ and $\phi$, respectively [see panel (b) and main text]. (b) Sketch of a one-dimensional quantum Ising spin chain with transverse magnetic field strength $\Omega$ and power-law interactions (exponent $\alpha$). Spins decay from their up-state $\mid\uparrow\rangle$ to the down-state $\mid\downarrow\rangle$ at a rate $\gamma$. (c) Unit sphere depicting spin rotation angles $\theta$ and $\phi$, for which the unitary $U(\theta,\phi)$ leads to an accelerated approach to the stationary state. The corresponding angles are indicated in red color.}
    \label{fig:schematics}
\end{figure}

This idea can be generalized to the dynamics of open quantum systems, as recently discussed in Ref. \cite{carollo2021}. Here it was shown that for any given pure initial state, $\rho_0$, relaxation to stationarity can be accelerated by the application of a preliminary unitary operation $U$, see Fig. \ref{fig:schematics}a. It was shown how this unitary needs to be constructed such that the new initial state $\bar{\rho}_0=U\rho_0U^\dagger$ arrives at stationarity exponentially faster. The idea underlying the construction of the unitary is that it should make the initial state ``orthogonal" to, i.e. not overlapping with, the dynamical mode associated with the slowest decay rate, which in fact corresponds to the spectral gap of the dynamical generator. For certain cases this unitary can be explicitly constructed. However, this construction may be not so useful in practical terms as one needs knowledge of the precise structure of some eigenmodes of the dynamical generator. The resulting unitary is typically rather contrived, i.e. it remains doubtful that it can be readily implemented in an experimental setting. 

In this work we investigate whether it is possible to achieve an exponential acceleration in the relaxation towards stationarity through the application of substantially simpler unitaries. To be specific, we focus on driven dissipative spin chains, shown in Fig. \ref{fig:schematics}b, which feature power-law interactions and single-spin decay. These systems model typical many-body settings realized on quantum simulators using trapped ions \cite{blatt2012} and neutral atoms \cite{browaeys2020} and are thus of direct relevance to current efforts in quantum computation and simulation. We show that already a global rotation --- solely dependent on the two angles $\theta$, $\phi$ which parametrize the unit sphere --- can be sufficient to observe a quantum Mpemba effect. In order to obtain a quantitative picture, we compute --- as a function of interaction strength and range --- the area on the unit sphere for which an exponentially accelerated approach to stationarity is achieved (see Fig. \ref{fig:schematics}c). We find that the Mpemba effect in Markovian open quantum systems is robust in the sense that it occurs with a simple unitary for a variety of angles in all studied parameter regimes. Moreover, our simplified approach also works in situations where it is not possible to analytically construct the ideal unitary with the approach put forward in Ref. \cite{carollo2021}. 

Our results may be of practical relevance for applications in quantum technology, e.g. for accelerating the dissipative preparation of entangled states and the processing speed of dissipative quantum computation \cite{verstraete2009} and pattern recognition \cite{fiorelli2020,marsh2021,carollo2021a,fiorelli2022}.

\section{Quantum Mpemba effect}
The open quantum systems we are considering in this work are described by a Markovian Lindblad Master equation \cite{breuer2002} which evolves the density matrix $\rho$ with the dynamical generator (Master operator) $W$:
\begin{eqnarray}
\dot{\rho}(t) &= &W [\rho(t)] \label{eq:master_operator}\\
&=& -i\left[H,\rho(t)\right]+\sum_k \left(L_k\rho(t) L_k^\dagger-\frac{1}{2}\left\{L_k^\dagger L_k,\rho(t)\right\}\right).\nonumber 
\end{eqnarray}
Here $H$ is the quantum Hamiltonian which governs the coherent dynamics, and the $L_k$ are the so-called jump operators which implement incoherent and dissipative processes \cite{breuer2002}.

The time-evolution of any initial state $\rho_0$ is given by
\begin{eqnarray}
\rho(t)&=&e^{W t} [\rho_0]=\rho_\mathrm{SS}+\sum_{k=2}^{D^2} \mathrm{tr}(\l_k \rho_0) r_k e^{\lambda_k t}. \label{eq:time_evolution}
\end{eqnarray}
Here, $D$ is the dimension of the system Hilbert space, and the $l_k$, $r_k$ and $\lambda_k$ are the left eigenmatrices, the right eigenmatrices and the eigenvalues of the dynamical generator, respectively:
\begin{eqnarray}
W [r_k] = \lambda_k r_k &\qquad,\qquad& W^\dagger [l_k] = \lambda_k l_k\, ,
\end{eqnarray}
where $W^\dagger$ is the dual (also called adjoint) Master operator which acts on observables rather than on states. 
The stationary state of the time-evolution, $\rho_\mathrm{SS}$, is the right eigenmatrix of the dynamical generator associated with the eigenvalue zero and has been taken out of the sum. The remaining eigenvalues $\lambda_k$ have a real part that is smaller or equal than zero. In the following we assume that the $\lambda_k$ are sorted in ascending order according to the modulus of their real part: $|\mathrm{Re}\left\{\lambda_{k+1}\right\}| \geq |\mathrm{Re}\left\{\lambda_{k}\right\}|$. In this case $\lambda_2$ corresponds to the spectral gap, and the negative inverse of its real part sets the longest timescale of the relaxation dynamics. This means, that for long times one has
\begin{eqnarray}
||\rho(t)-\rho_\mathrm{SS}|| \sim \exp\left(\mathrm{Re}\left\{\lambda_2\right\} t\right),
\end{eqnarray}
where $||\cdot||$ is a suitably chosen norm. From the expansion (\ref{eq:time_evolution}) it then follows that if there exists a unitary $U$ which eliminates the overlap of the initial state with this slowest decaying mode, i.e.
\begin{eqnarray}
\mathrm{tr}(\l_2 U\rho_0U^\dagger) = 0, \label{eq:zero_overlap}
\end{eqnarray}
the approach to the stationary state is exponentially accelerated. The new timescale over which the stationary state is approached is given by the inverse of the negative real part of $\lambda_3$: $\tau_3=-1/\mathrm{Re}(\lambda_3)$  (see sketch in Fig. \ref{fig:schematics}a)
\begin{eqnarray}
||\rho(t)-\rho_\mathrm{SS}|| \sim \exp\left(\mathrm{Re}\left\{\lambda_3\right\} t\right).
\end{eqnarray}

In Ref. \cite{carollo2021} it was shown that the unitary $U$ accomplishing condition (\ref{eq:zero_overlap}) can be explicitly constructed for a pure initial state --- which we write as $\rho_0=\dyad{0}$ --- provided that the eigenvalue $\lambda_2$, which corresponds to the slowest decaying mode, is real and non-degenerate. In this case the corresponding left-hand eigenmatrix, $l_2$, is hermitian and can be spectrally decomposed as
\begin{eqnarray}
l_2=\sum_{m=1}^D \alpha_m \dyad{\phi_m}, \label{eq:l2_expansion}
\end{eqnarray}
with the eigenvalues $\alpha_m\,\in\,\mathbb{R}$ and the orthonormal set of eigenstates $\left\{\mid\!\phi_m\rangle\right\}$. If one of the eigenvalues is zero --- without loss of generality we assume $\alpha_1=0$ --- then condition (\ref{eq:zero_overlap}) can be met by a unitary that rotates the initial state $\ket{0}$ onto the corresponding eigenstate $\ket{\phi_1}$, such that $\left|\!\mel{\phi_1}{U}{0}\right|=1$.

In general, the expansion (\ref{eq:l2_expansion}) does not contain a zero eigenvalue. A unitary, which satisfies condition (\ref{eq:zero_overlap}) can nevertheless be constructed. Underlying this construction is the observation that among the eigenvalues $\alpha_m$ some have to be positive and some negative. This results from the fact that the overlap between $l_2$ and the stationary state vanishes, i.e.,
\begin{eqnarray}
0=\mathrm{tr}(l_2 \rho_\mathrm{SS})=\sum_{m=1}^D \alpha_m \langle\phi_m |\rho_\mathrm{SS}|\phi_m \rangle.
\end{eqnarray}
Given that the stationary state is a positive operator, i.e. $\langle\phi_m |\rho_\mathrm{SS}|\phi_m \rangle \geq 0$, it follows that there are both positive and negative $\alpha_m$. Without loss of generality we assume that $\alpha_1<0$ and $\alpha_2>0$. Introducing the unitary
\begin{eqnarray}
U(s)=\exp\left[-is\left(|\phi_1\rangle\langle \phi_2|+|\phi_2\rangle\langle \phi_1|\right)\right] R,
\label{eq:ideal_unitary}
\end{eqnarray}
with $R$ being also a unitary operator such that  $R\ket{0}=\ket{\phi_1}$, one can show that \cite{carollo2021}
\begin{eqnarray}
\mathrm{tr}(\l_2 U(s)\rho_0U^\dagger(s))=\alpha_1 \cos^2(s)+\alpha_2 \sin^2(s),
\end{eqnarray}
and thus by choosing
\begin{eqnarray}
s=\arctan\left(\sqrt{-\frac{\alpha_1}{\alpha_2}}\right)
\end{eqnarray}
condition (\ref{eq:zero_overlap}) is met.

Unfortunately, this way of constructing the  unitary operator accelerating the relaxation to stationarity is not justified in other situations. For example, a scenario that is often encountered is one where the eigenvalues of the lowest excited modes of the Master operator form a complex conjugate pair:
\begin{eqnarray}
\rho(t)&=&\rho_\mathrm{SS}+\mathrm{tr}(\bar{l}_2 \rho_0)\, \bar{r}_2 e^{\lambda_2 t}+\mathrm{tr}(\bar{l}_2^\dagger \rho_0)\, \bar{r}_2^\dagger e^{\lambda^*_2 t} + ...\label{eq:conjugate_pair}
\end{eqnarray}
To achieve acceleration one would seek a unitary $U$ that accomplishes
\begin{eqnarray}
\mathrm{tr}(\bar{l}_2 U \rho_0U^\dagger) = [\mathrm{tr}(\bar{l}_2^\dagger U\rho_0U^\dagger)]^*= 0.
\end{eqnarray}

However, due to the fact that $\bar{l}_2$ is not hermitian in these cases, a decomposition like the one of Eq. (\ref{eq:l2_expansion}) is not possible and the previously presented route for constructing $U$ cannot be pursued.

\section{Accelerated approach to stationarity via simple unitaries}
While the unitary transformation (\ref{eq:ideal_unitary}) can be in theory constructed, it will be in practice challenging to implement given that it requires precise knowledge of the mode $l_2$. Moreover, when the eigenvalues of the slowest decaying modes form a complex conjugate pair the theory of the previous section is not applicable. It is therefore of relevance to understand whether simpler unitaries exist, that also may lead to the desired acceleration towards stationarity. This appears possible given that the mere requirement for achieving acceleration is a vanishing overlap between the transformed initial state and the slowest decaying mode. For sufficiently large state spaces, it is reasonable to expected that this should be possible to achieve with a simple unitary. 

To investigate this we consider for concreteness a one-dimensional spin system, composed by $N$ spins, whose coherent evolution is governed by the Hamiltonian (see Fig. \ref{fig:schematics}a) 
\begin{eqnarray}
H=\Omega \sum_{k=1}^N \sigma_k^x + V \sum_{k<m}^N \frac{\sigma_k^z\sigma_m^z}{|k-m|^\alpha},
\end{eqnarray}
where the $\sigma_k^\nu$ ($\nu=x,y,z$) are the usual Pauli matrices. The parameter $\Omega$ is the transverse field strength, $V$ is the interaction strength and $\alpha$ is the exponent that controls the interaction range \cite{marcuzzi2014}. Dissipation is governed by the jump operators
\begin{eqnarray}
L_k=\sqrt{\gamma} \sigma_k^- = \frac{\sqrt{\gamma}}{2}(\sigma_k^x-i \sigma_k^y),
\end{eqnarray}
which model single-spin decay from the up-state $\mid\uparrow\rangle$ to the down-state $\mid\downarrow\rangle$ at rate $\gamma$. Related models have been theoretically studied in the past \cite{ates2012,kessler2012,weimer2015,jin2018,paz2021}, and instances of these systems can be experimentally realized in systems (quantum simulators) consisting of trapped ions \cite{kim2011,blatt2012} and Rydberg atoms \cite{malossi2014,letscher2017,browaeys2020}.
\begin{figure}
    \flushleft
    \includegraphics[width=.9\columnwidth]{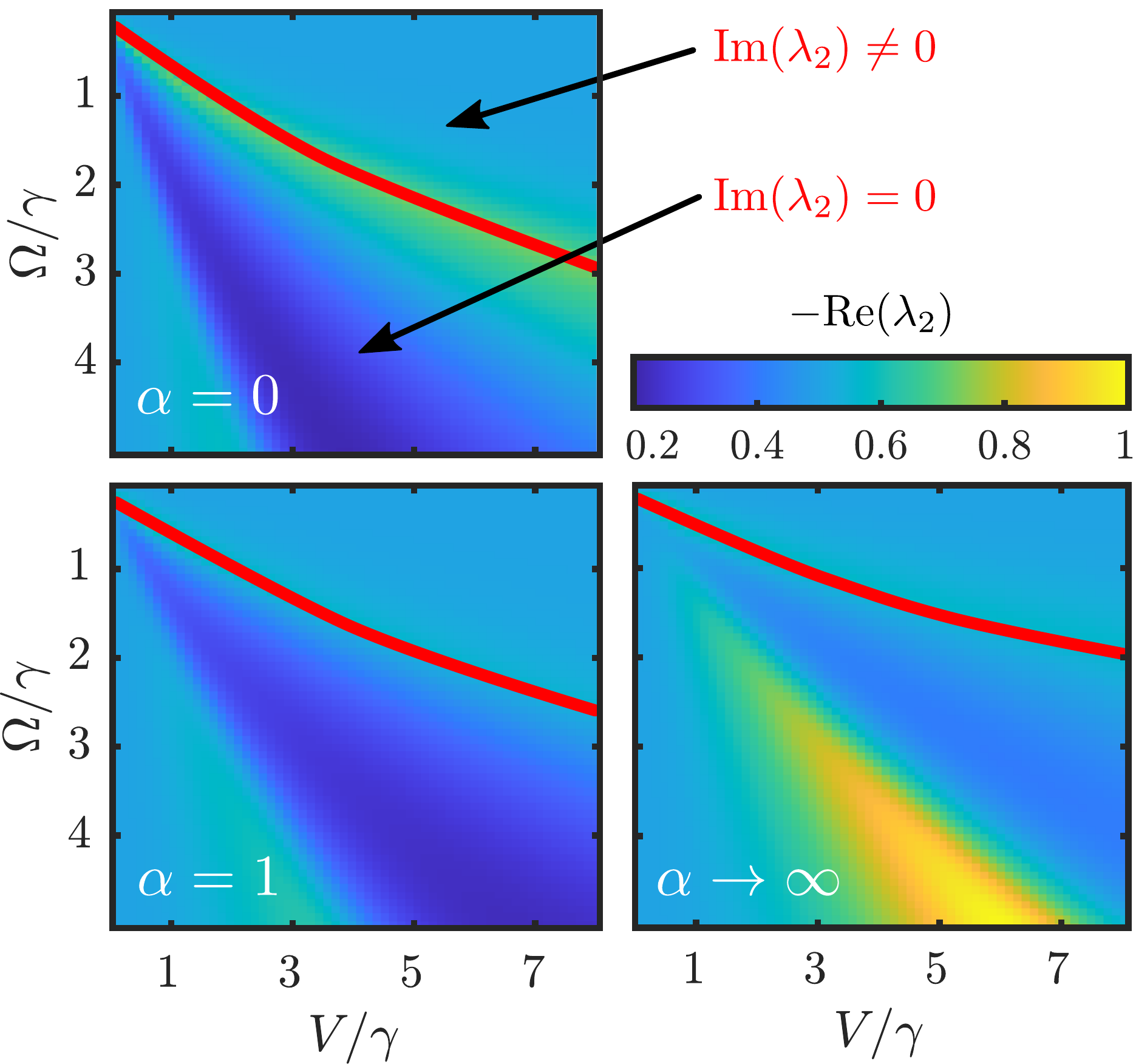}
    \caption{\textbf{First excited mode of the Master operator.} The density plot shows the negative
    real part of the first non-zero eigenvalue $\lambda_2$ of the Master operator $W$ [see Eq. (\ref{eq:master_operator})] in the $\Omega-V$-plane for three different values of the power-law exponent $\alpha$. The red line delimits the region where $\lambda_2$ has a zero and a non-zero imaginary part. The latter case corresponds to the situation described by Eq. (\ref{eq:conjugate_pair}). Note, that for obtaining the data presented here we have chosen the lowest excited eigenvalue of $W$ which lies in the same (permutation invariant) symmetry subspace as the initial state (\ref{eq:initial_state}). The data shown is for a system size of $N=5$ and open boundary conditions.}
    \label{fig:lambda2}
\end{figure}

In order to obtain first insights into the model we show in Fig. \ref{fig:lambda2} the negative real part of the eigenvalues  $\lambda_2$ of the Master operator [see spectral expansion given by Eq. (\ref{eq:time_evolution})]. Moreover, we indicate where $\lambda_2$ possesses a non-zero imaginary part and the time-evolved state is thus of the form (\ref{eq:conjugate_pair}). As can be seen in the figure, this is indeed a relevant case, which occurs across a wide region of parameters.

For the purpose of this work we consider a pure initial state in which all spins occupy the down-state:
\begin{eqnarray}
\rho_0=\mid\downarrow\downarrow ... \downarrow \rangle \langle \downarrow\downarrow ... \downarrow\mid. \label{eq:initial_state}
\end{eqnarray}
The unitary that we apply to this initial state has the form
\begin{eqnarray}
U(\theta,\phi)=\prod_{k=1}^N U_S^{(k)}(\theta,\phi), \label{eq:unitary}
\end{eqnarray}
with 
\begin{eqnarray}
U_S^{(k)}(\theta,\phi) = \exp\left(\frac{i}{2} \phi \sigma^z_k \right)\, \exp\left(\frac{i}{2} \theta \sigma^y_k \right). 
\end{eqnarray}
This unitary, which acts globally on all spins simultaneously, rotates them first by an angle $\theta\, \in\, [0,\pi]$ (polar angle) about the $y$-axis and subsequently by an angle $\phi\, \in\,  [0,2\pi)$ (azimuthal angle) around the $z$-axis. On quantum simulator platforms such operation can be readily realized by suitably timed laser or microwave pulses and is therefore significantly easier to implement than the ideal unitary (\ref{eq:ideal_unitary}).

\begin{figure}
    \flushleft
    \includegraphics[width=\columnwidth]{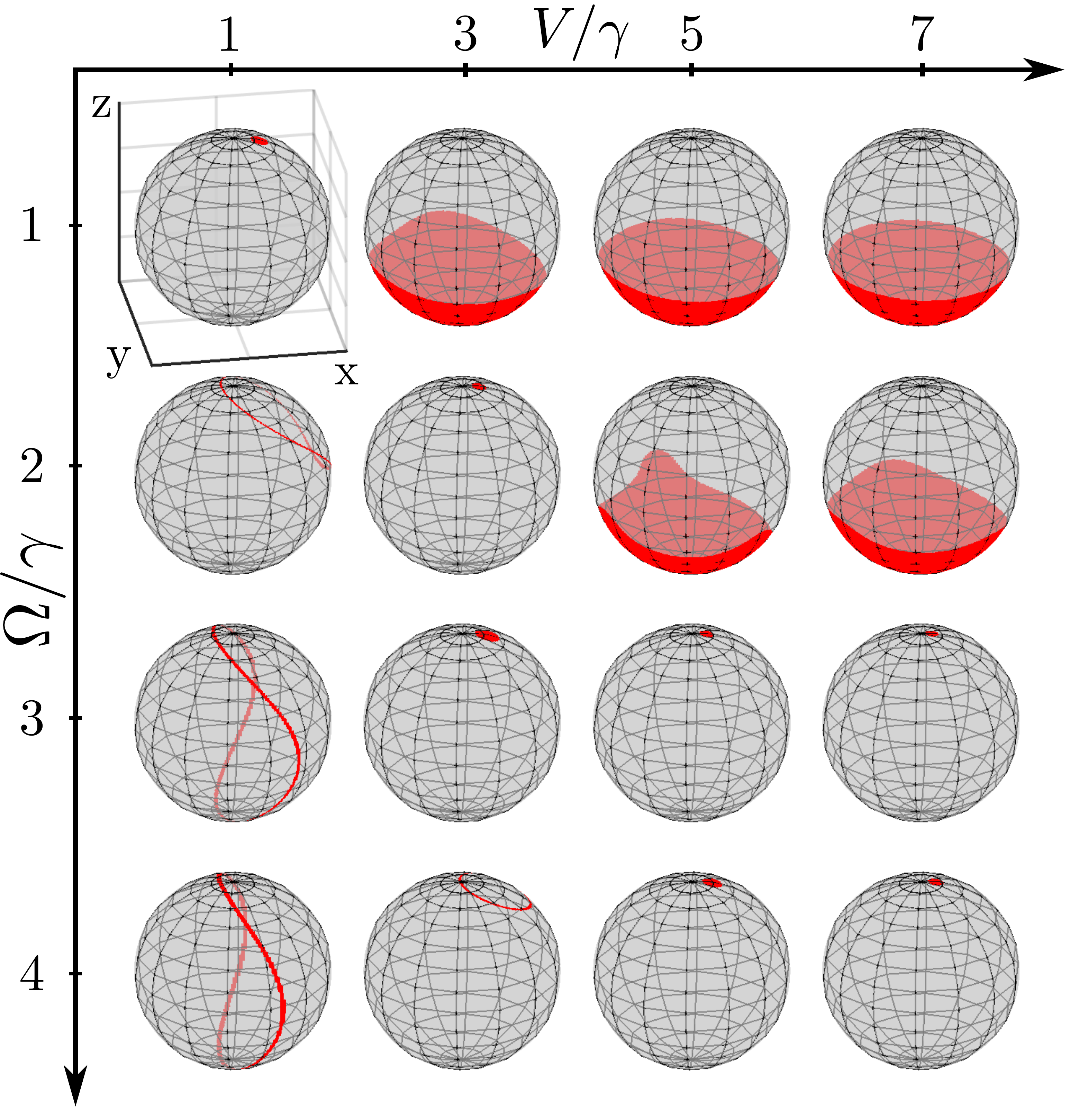}
    \caption{\textbf{Rotation angles which lead to an accelerated approach to stationarity.} The application of the unitary (\ref{eq:unitary}) to the initial state (\ref{eq:initial_state}) can reduce its overlap with the slowest decaying mode of the Master operator (\ref{eq:master_operator}). Angles for which this overlap, $\chi(\theta,\phi)$, is smaller than $\epsilon=10^{-2}$ --- see Eq. (\ref{eq:overlap}) --- are displayed in red. The abrupt change in the size of the red region for large $V$ is due to a crossing of the eigenvalues of the Master operator, which can also be observed in Fig. \ref{fig:lambda2}. The data shown is for $N=5$ and power-law exponent $\alpha=0$. Note, that if one were to decrease the threshold value $\epsilon$ the width of the red ribbon-like lines (see small values of $V$) would shrink and the extended red areas (see large values of $V$) would contract.}
    \label{fig:mosaik}
\end{figure}
The goal is now to determine the angles $\theta$ and $\phi$ for which acceleration is achieved. To accomplish this we use the criterion
\begin{eqnarray}
\chi(\theta,\phi)=|\mathrm{tr}(l_2 U(\theta,\phi)\rho_0 U^\dagger(\theta,\phi))| \leq \epsilon, \label{eq:overlap}
\end{eqnarray}
where $\epsilon$ is a threshold, which we set to $10^{-2}$. In Fig. \ref{fig:mosaik} we depict these angles for various parameter regimes of the considered spin system. We find that it is indeed generally possible to find rotation angles $(\theta,\phi)$ which reduce the overlap of the initial state and the lowest decaying mode below the chosen threshold. In fact, there are entire areas on the unit sphere for which acceleration of the approach to stationarity is achieved. As can be seen in Fig. \ref{fig:mosaik} --- where all the rotation angles leading to acceleration are marked in red --- the size of this area on the unit sphere may depend strongly on the values of the parameters $\Omega$ and $V$. On the one hand, when both $\Omega$ and $V$ are simultaneously large or small only a small rotation ($\theta \ll \pi$) away from the initial state ($\theta=0$, $\phi=0$) suffices to speed up the approach to stationarity. On the other hand, one observes that for small values of $\Omega$ and large values of $V$ the area on the unit sphere for which acceleration is achieved becomes large. This drastic quantitative change is due to the fact that the real parts of the eigenvalues corresponding to the slowest decaying mode of the Master operator cross, and therefore the character of the slowest decaying mode is changing. In Fig. \ref{fig:lambda2} this eigenvalue crossing also manifests, e.g., through the appearance of an imaginary part in the eigenvalue corresponding to the slowest decaying mode.

\begin{figure}
    \flushleft
    \includegraphics[width=0.9\columnwidth]{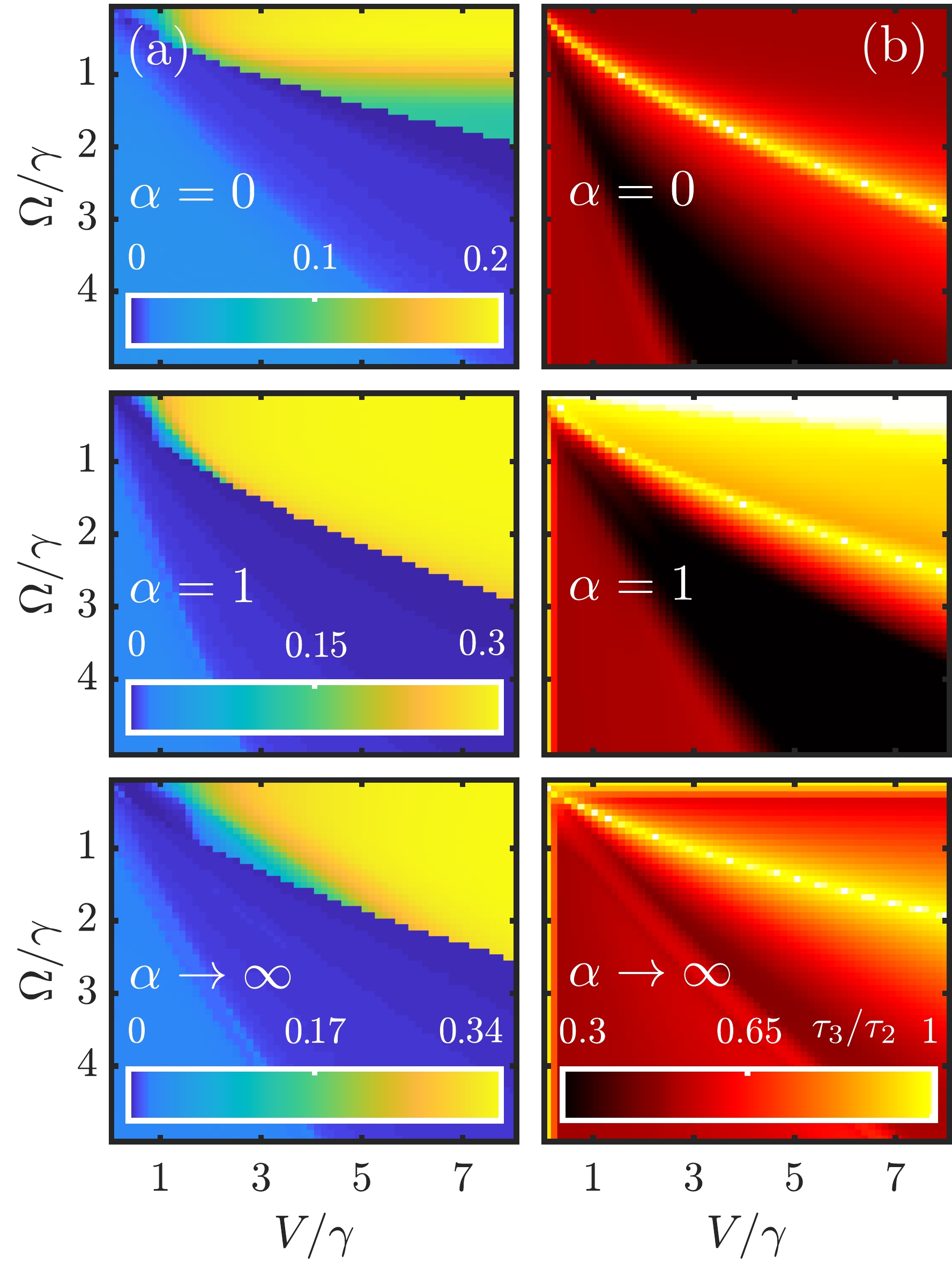}
    \caption{\textbf{Relative size of parameter region in which acceleration is achieved and gain in relaxation time.}
    (a) Relative area (\ref{eq:area}) on the unit sphere for which the overlap of the initial state with the slowest decaying mode is smaller than $\epsilon=10^{-2}$. This area corresponds to the regions marked in red in Fig. \ref{fig:mosaik}. (b) Ratio of the timescales $\tau_3=-1/\mathrm{Re}(\lambda_3)$ and $\tau_2=-1/\mathrm{Re}(\lambda_2)$ which quantifies the minimal reduction in relaxation timescale which can be achieved by removing the overlap between the initial state and the slowest relaxing mode of the Master operator \eqref{eq:master_operator}. Lower values mean larger acceleration. Note, that for obtaining the data presented here we have chosen the modes of the Master operator $W$ that lie in the same (permutation invariant) symmetry subspace as the initial state (\ref{eq:initial_state}). The data shown in the panels is for a system size of $N=5$ (open boundary conditions) and various values of the power-law exponent $\alpha$.}
    \label{fig:acceleration_area}
\end{figure}
In the following we investigate the relative area on the unit sphere spanned by the angles $\theta$ and $\phi$, where acceleration is achieved, i.e. the size of the red areas in Fig. \ref{fig:mosaik}. This quantity is defined as
\begin{eqnarray}
A=\frac{1}{4\pi}\int_0^{2\pi}\!\!d\phi\, \int_0^{\pi}d\theta \sin\theta\, \Theta\!\left[\epsilon-\chi(\theta,\phi)\right],\label{eq:area}
\end{eqnarray}
where $\Theta[x]$ is the Heaviside step-function and $\epsilon$ is the threshold parameter introduced above. The corresponding data is shown in Fig. \ref{fig:acceleration_area}(a) for a system of $N=5$ spins \footnote{We have also studied systems composed of $3$ and $4$ spins, which showed comparable results.}. Throughout the entire $\Omega-V$-plane we find $A$ to be non-zero, which means that it is always possible to find a global unitary (\ref{eq:unitary}) that accomplishes a speed-up of relaxation. However, the size of the area may differ considerably. In particular for large values of $V$ the area is significantly extended which might offer some robustness with respect to variations in the angles $\theta$ and $\phi$ that could be relevant for experiments. This change from small to large area is not gradual, but appears at a critical value of $V$, which increases with $\Omega$, regardless of the value of $\alpha$, i.e. the exponent describing the power-law decay of the interactions. This abrupt change might be a (finite-size) signature of a dissipative phase transition in the stationary state of the dissipative spin chain, which has been studied previously in several works \cite{lee2012,ates2012,marcuzzi2014,weimer2015,ates2012}.

As a final point of investigation, we analyze the achievable acceleration. To this end we compare the timescales corresponding to the slowest and second slowest decaying mode, which are given by $\tau_2=-1/\mathrm{Re}(\lambda_2)$ and $\tau_3=-1/\mathrm{Re}(\lambda_3)$, respectively. In Fig. \ref{fig:acceleration_area}(b) we show the ratio of the two timescales $\tau_3/\tau_2$, i.e. the smaller this quantity the higher the achievable acceleration. As can be seen from the data, the relaxation timescale can be reduced up to a factor of three, depending on the specific values of $\Omega$ and $V$. In the region where the abrupt transition is visible in panel (a) hardly any acceleration is possible. This means that the eigenvalues of the slowest and second slowest decaying mode become degenerate, and corroborates the picture of the onset of a phase transition which is typically associated with collapsing eigenvalues of the dynamical generator. 

\section{Conclusions and outlook}
We have shown --- using an open transverse field Ising model as exemplary case --- that the approach to stationarity of an open quantum system can be accelerated by a unitary rotation of the initial state. The underlying idea of this so-called Mpemba effect is that such rotation renders the initial state orthogonal to the slowest decaying mode of the Master operator. Given the high-dimensional state space of a many-body system, it appears plausible that it should not be too challenging to achieve this. Indeed, even for the global unitary considered in our work it was always possible to find spin rotation angles that in principle yield an exponential speed-up of the relaxation dynamics. This is an encouraging finding, given that this type of unitary is simple to implement on current quantum computing and simulation platforms. 

There are, nevertheless, a number of open questions which give room for further studies. For example, the current investigation focuses on the case of a pure initial state and an extension of the investigation of the Mpemba effect to the case of mixed states is highly desirable. Moreover, one may ask whether and under what conditions it is possible to render the initial state orthogonal to a large set of slowly-decaying modes. This may become particularly important in many-body systems in the vicinity of phase transitions, where the spectrum is dense and many eigenstates of the Master operator acquire a real part close to zero. On the one hand these questions can be explored numerically in certain model systems, such as the spin chains considered here. On the other hand, it would be interesting to understand under what circumstance it is --- at least theoretically --- possible to explicitly construct unitaries or, more generally, quantum maps that remove the overlap of a given initial state with a set of slowly decaying modes.

\acknowledgments
We acknowledge discussions with M. Hennrich, A. Lasanta, D. Manzano, O. Morsch, Ö. E. Müstecaplıoğlu and R. Sanchez. The research leading to these results has received funding from the “Wissenschaftler-R\"uckkehrprogramm GSO/CZS” of the Carl-Zeiss-Stiftung and the German Scholars Organization e.V., through the Deutsche Forschungsgemeinsschaft (DFG, German Research Foundation) under Projects No. 435696605 and 449905436, as well as from the Baden-W\"urttemberg Stiftung through Project No.~BWST\_ISF2019-23.
\bibliography{biblio}
\end{document}